\documentclass[reprint,amsmath,amssymb,aps,]{revtex4-1}
\usepackage{graphicx}
\usepackage{amsmath}
\usepackage{subfigure}
\usepackage{mathtools} 
\usepackage{color}
\usepackage{overpic}
\usepackage[export]{adjustbox}
\usepackage{mwe, tikz}
\usepackage{rotating}
\usepackage{latexsym}
\usepackage{csquotes}
\usepackage{tikz} 
\usetikzlibrary{calc} 
\usepackage{pict2e}
\usepackage{pgfplots}
\usepackage{hyperref}
\usepackage{filecontents}
\usepackage{natbib}

\begin{document}
\title{Correlation and shear bands in a plastically deformed granular medium}
\author{Kamran Karimi}
\affiliation{Universit\'e Grenoble Alpes, CNRS, ISTerre, 38041 Grenoble cedex 9, France}
\author{Jean-Louis Barrat}
\affiliation{Universit\'e Grenoble Alpes, CNRS, LIPHY, F-38000 Grenoble, France}
\begin{abstract}
Recent experiments (Le Bouil \emph{et al.}, \emph{Phys. Rev. Lett.}, 2014, \textbf{112}, 246001 \cite{le2014emergence}) have analyzed the statistics of local deformation in a granular solid undergoing plastic deformation. 
Experiments report strongly anisotropic correlation between events, with a characteristic angle that was interpreted using elasticity theory and the concept of Eshelby transformations with dilation; interestingly, the shear bands that characterize macroscopic failure occur at an angle that is different from the one observed in microscopic correlations.
Here, we interpret this behavior using a mesoscale elastoplastic model of solid flow that incorporates a local Mohr-Coulomb failure criterion. 
We show that the angle observed in the microscopic correlations can be understood by combining the elastic interactions associated with Eshelby transformation with the local failure criterion. 
At large strains, we also induce permanent shear bands at an angle that is different from the one observed in the correlation pattern. 
We interpret this angle as the one that leads to the maximal instability of slip lines.
\end{abstract}
\maketitle
\section{Introduction}
Plasticity is an important mechanical property in a wide variety of amorphous systems such as dense colloidal glasses, foams, emulsions, and fine-grained granular packings.
It is formally defined as intense unrecoverable (shear) deformations that the material undergoes beyond its elastic limit without any crushing or crumbling.
This phenomenon has been linked to the so-called \emph{yielding transition} \cite{LinPNAS2014, LinEPL2014, FerreroPRL2014, DenisovNcomm2016} with some \emph{universal} features associated with it.  Universality emerges in spite of the diversity in disordered solids -- in terms of their scales, microscopic constituents, or interactions, suggesting common underlying mechanisms.

A commonly accepted picture that supports this universal character is that the \emph{bulk} plastic response emerges from a collective  dynamics that is not specific  to the particle \footnote{Here particles may represent grains in granular materials, bubbles in concentrated foams, or colloids in suspensions.}, but rather results from interactions mediated by the universal laws of linear elasticity.
This emergent dynamics is characterized by \emph{ plastic events} or \emph{ shear transformations} that are localized in space and time, but have long-range  (compared to the size of rearranging zones) elastic-type consequences \cite{picard2004elastic}.
In systems in which thermal fluctuations are irrelevant (which will be the case of the granular systems considered in this work), rearrangements are initially activated by external deformation, but further instability may be triggered and propagated due to non-local interactions.

In this framework, propagation of plasticity is a dynamical process which, once the characteristics of the shear transformations and the  elastic properties of the medium are known,  depends only  on the dissipation mechanism \cite{karimi2017inertia, SalernoPRL2012}.    
Near the yielding transition, the so-called \emph{avalanche} dynamics may emerge in which the activation process takes place by sequentially forming  clusters of all scales.
In this regard, plastic yielding may be thought as a true second-order phase transition with unique characteristics such as diverging length and/or timescales and power-law distributions of avalanche sizes \cite{BakPRL1987, zapperi1997plasticity,ParisiPNAS2017}.

Due to structural heterogeneities and stochastic aspects of the shear transformation dynamics, the cascades of  events described as avalanches are usually 
organized in a highly intermittent manner in space and  time.  Intermittency makes them quite distinguishable from the much longer lived localization patterns that emerge upon ultimate failure and are often known as \emph{shear bands}, i.e. narrow   linear (in 2$d$) or planar  (in 3$d$) structures along which plastic activity accumulates while the bulk of the material of the system remains undeformed.  Still, it is expected that pre-failure collective dynamics must have a strong connection with  the formation of permanent shear bands. In fact, it has been shown in 
several works that elasticity-based polar features  characterize the structure of  correlations  between plastic bursts \cite{puosi2014time, jensen2014local,karimi2015elasticity}, which occur preferentially at $45^\circ$  for a volume conserving plastic event. This preferential direction (with respect to the principal axis of the local shear event)  is also the one along which a shear band should form in a system deformed at constant volume (in particular incompressible), as it corresponds to the direction of maximum macroscopic shear stress.

However, the  symmetry that Le Bouil \emph{et al.} \cite{le2014emergence} (see also Ref.~\cite{le2014biaxial} for the experimental setup) observed  in correlation patterns of the plastic activity  seems to be at odds with the morphology of the fully formed band they also observed in strongly deformed granular media. 
While the latter was formally described as reflecting the Mohr-Coulomb macroscopic failure angle, the former revealed a distinct orientation.   Moreover, both angles differ from the ``canonical'' $45^\circ$.  In order to rationalize the preferential direction observed in correlations, the authors proposed to take into account the possibility of a local dilation in the plastic event, a feature that was already noted to affect the preferential directions in collective bursts plastic activity \cite{AshwinPRE2013}.
The authors proposed a phase co-existence scenario in which recurring mini bursts persist up until failure, which  may be interpreted as a signature of discontinuous first-order or spinodal  transition. However, this picture does not lead to a specific prediction for the angle of failure in the macroscopic system.

In this work, our aim is to provide, based on minimal ingredients, a scenario that explains simultaneously the deviation from the ``canonical'' $45^\circ$ direction for the different observables, and the fact that different directions are observed for the correlations in intermittent activity and in permanent shear localization. Our analysis will be based on the hypothesis that, while the interactions between plastic events are mediated by the universal laws of linear elasticity, the triggering of events depends on a local failure criterion that is completely independent of these laws, and can be quite arbitrary. As a result, the correlations in plastic activity and its propagation involve a compromise between directions favored by the elastic interactions and those favored by the local failure criteria. The results that emerge from this compromise are nontrivial, in agreement with the experimental observations. 

The minimal ingredients used  in our analysis are the description of local plastic events as transformations of Eshelby-type \cite{eshelby1957determination}, combined with the use of a local Mohr-Coulomb criterion that introduces a pressure sensitivity in the propagation of plasticity. It can be seen as an extension to the case of Mohr-Coulomb failure of works that involve permanent damage as a cause of strain localization  \cite{amitrano1999diffuse, vandembroucq2011mechanical,BerthierJMPS2017}. Remarkably, however, no permanent damage is required for inducing coexistence between transient micro events and fully developed shearing bands. 

%
The organization of the paper is the following.
In Sec.~\ref{sec:mcTheory} we combine  the  hypothesis of Mohr-Coulomb local failure with the stress redistribution prescribed by Eshelby's elastic theory, and discuss hypothetically the consequences on the correlation patterns and macroscopic failure angle. In
Sec.~\ref{sec:numMod} we describe the numerical model that incorporates these basic ingredients.
The numerical results and conclusions are given in Sec.~\ref{sec:NumRes} and Sec.~\ref{sec:cnlsn}, respectively.

\section{The Mohr-Coulomb Failure Criterion}\label{sec:mcTheory}
Previous studies within the framework of mesoscopic elasto-plasticity \cite{martens2012spontaneous, lin2014density} emphasized the role of elastic kernels in the yielding transition.
The local yielding rule however, coupled with long-range elasticity, must have a strong relevance on the structure of correlations and macroscopic failure.
In this section, we propose a simple theory taking into account these effects at both \emph{local} and \emph{macroscopic} scales.

\subsection{Meso-scale Failure Criterion and Elastic Stress Redistribution}
We consider an infinite elastic matrix  characterized by its bulk modulus $K$ and shear modulus $\mu$ and an embedded inclusion going through a shear transformation $\epsilon^\text{tz}_{\alpha\beta} = \epsilon^*a^d(\delta_{x\alpha}\delta_{x\beta}-\delta_{y\alpha}\delta_{y\beta})$ with an amplitude $\epsilon^*$ and microscopic volume $a^d$. 
Here $\delta_{\alpha\beta}$ is a Kronecker delta and we focus on the two dimensional case $d=2$.

The stress tensor at a material point can be expressed as 
$\sigma_{\alpha\beta}=-p\delta_{\alpha\beta}+
\sigma(\delta_{\alpha x}\delta_{\beta x}-\delta_{\alpha y}\delta_{\beta y})+
\sigma_{xy}(\delta_{\alpha x}\delta_{\beta y}+\delta_{\alpha y}\delta_{\beta x})$ 
in terms of the pressure $p$ and area preserving axial shear $\sigma$ and diagonal shear $\sigma_{xy}$.
The far-field \emph{perturbation} fields given by Eshelby's solutions at a point with polar coordinates $r,\theta$ read \cite{eshelby1957determination}
\begin{eqnarray}
	\delta p &=& \frac{2\mu\epsilon^*}{1+\frac{\mu}{K}} (\frac{a}{r})^d~\text{cos}~2\theta,  \nonumber \\
	\delta\sigma &=& -\frac{2\mu\epsilon^*}{1+\frac{\mu}{K}} (\frac{a}{r})^d~\text{cos}~4\theta, \nonumber \\
	\delta\sigma_{xy} &=& -\frac{2\mu\epsilon^*}{1+\frac{\mu}{K}} (\frac{a}{r})^d~\text{sin}~4\theta,
\end{eqnarray}
for $r\gg a$. 
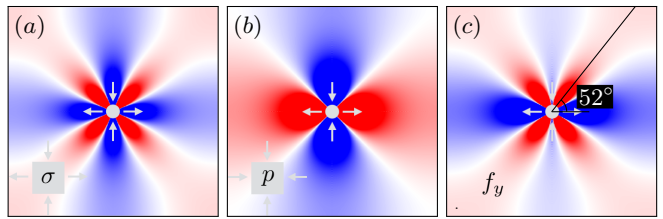
\begin{figure}
	\begin{center}
		\begin{overpic}[width=0.155\textwidth,frame]{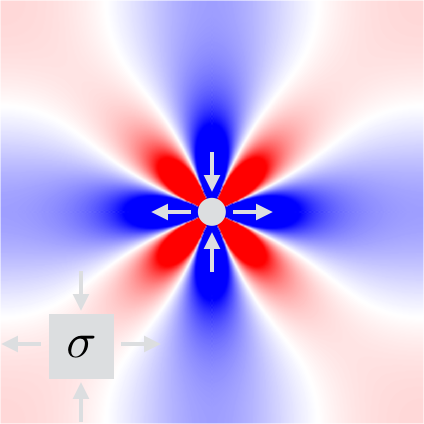}
			\put (3,87) {$\small(a)$}
		\end{overpic}
		\begin{overpic}[width=0.155\textwidth,frame]{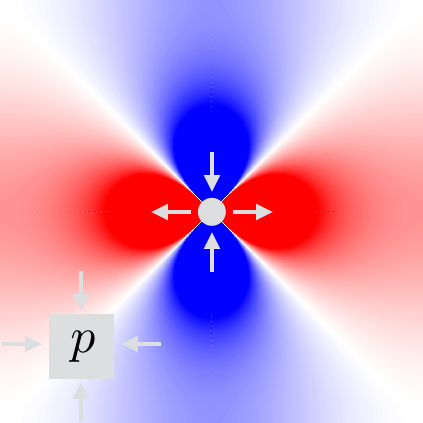}
			\put (3,87) {$\small(b)$}
		\end{overpic}
		\begin{overpic}[width=0.155\textwidth,frame]{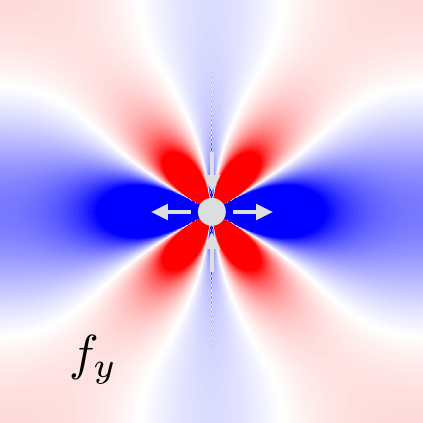}
			\put (3,87) {$\small(c)$}
			\begin{tikzpicture} 
				\coordinate (a) at (0,0); 
				\node[] at (a) {\tiny.};
				\coordinate (center) at (1.26,1.28); 
				\draw (center) -- ( $ (center) + 1.11*(1,1.259) $); 
				\draw (center) -- ( $ (center) + (0.5,0) $); 
				\draw ($(center)+(.2,0)$) arc(0:51.548:2mm); 
				\draw[fill=black] ($(center)+(.34,.05)$) rectangle ($(center)+(.34,.05)+(0.47,0.3)$); 
				\node[text=white] at ($(center)+(.6,.21)$) {\small{$52^\circ$}}; 
			\end{tikzpicture}
		\end{overpic}
	\caption{Far-field redistribution of a) axial shear $\sigma$ b) pressure $p$ c) yield function $f_y$ in an elastic matrix with $\frac{\mu}{K}=\frac{1}{2}$ in view of a shear transformation zone. Blue and red denote a decrease and increase in local stresses, respectively. The direction of maximal change in $f_y$ is marked in (c) at $\phi=65^{\circ}$.}
	\label{fig:distortionStz}
	\end{center}
\end{figure}
The power-law decay with distance $r$ marks the non-local long-range nature of the disturbance that also has distinct angular symmetries in terms of $\theta$. 
Note the four-fold symmetry in $\sigma$ which is contrasted by the bi-polar structure in $p$ as illustrated in Fig.~\ref{fig:distortionStz}(a) and (b).
The negative (blue) and positive (red) lobes represent regions with decreasing and increasing stresses in space, respectively. 

If at a point within the bulk the shear stress on any plane becomes equal to the shear \emph{strength}, failure will occur at that point. In the following, we will assume that the local 
shear strength can be expressed as a linear function of the confining pressure \cite{craig2013soil}, in accordance with the Mohr-Coulomb concept.  
Therefore, the distance to failure at any point can be expressed by the yield function $f_y$:
\begin{equation}
f_y = |\tau_m|-(p~\text{sin}~\phi+c~\text{cos}~\phi).
\end{equation}
Here $c$ and $\phi$ are the cohesion and the internal angle of  friction respectively, and $\tau_m = \sqrt{\sigma^2+\sigma_{xy}^2}$.

Assuming $\sigma_{xy}=0$, the far-field perturbations in $f_y$ to leading order following a shear transformation taking place at the origin  is given by
\begin{equation}
	\delta f_y=\delta\sigma-\delta p~\text{sin}~\phi.
\end{equation}
Obviously, failure is likely to localize into the overlapped sectors between positive lobes of $\delta\sigma$ in Fig.~\ref{fig:distortionStz}(a) and negative lobes of $\delta p$ in Fig.~\ref{fig:distortionStz}(b).  
Fig.~\ref{fig:distortionStz}(c) displays perturbations in $f_y$ for a value $\phi=65^{\circ}$ of the local friction. Near the shear transformation site, the pattern  retains the quadrupolar shape  with its positive lobes slightly tilted upwards toward regions with decreasing pressure.
Note that a positive $\delta f_y$ means that material points are pushed toward failure threshold. 
The direction of the maximum in the positive lobes (denoted by $\theta_\text{max}$ hereafter) will depend on the internal friction $\phi$ and is given by
\begin{equation}
	\label{eq:thetaMaxDist}
	\frac{\partial}{\partial \theta}\delta f_y|_{\theta=\theta_\text{max}}=0:\text{cos}~2\theta_\text{max} = -\frac{1}{4}\text{sin}~\phi,
\end{equation}
or $\theta_\text{max}\approx 45^{\circ}+\frac{\phi}{8}$ for low values of $\phi$. 
In the absence of friction, \emph{i.e.} $\phi=0$, we recover $\theta_\text{max}=45^{\circ}$ implying that the change in $f_y$ takes its maximal value along the direction of maximum change in shear stress. 

In addition to distortion, we now suppose that transformation zones may undergo dilation as well, \emph{i.e.} $\epsilon^\text{tz}_{\alpha\beta} = \frac{1}{2}\epsilon_v^*a^d\delta_{\alpha\beta}$ with dilatancy $\epsilon_v^*$. 
This will make an anisotropic contribution with the two-fold symmetry to the shear stress $\delta\sigma = -\frac{\mu\epsilon_v^*}{1+\frac{\mu}{K}} (\frac{a}{r})^d~\text{cos}~2\theta$. 
Combining the effects of the local dilation and shear yields
\begin{equation}
	\label{eq:thetaMaxDilt}
	\text{cos}~2\theta_\text{max} = -\frac{1}{4}(\frac{\epsilon_v^*}{2\epsilon^*}+\text{sin}~\phi).
\end{equation}
In comparison with Eq.~\ref{eq:thetaMaxDist}, $\theta_\text{max}$ in the above equation has an extra ingredient that includes $\frac{\epsilon_v^*}{2\epsilon^*}$ the ratio between volumetric and shear strains. 
In the limit $\frac{\epsilon_v^*}{2\epsilon^*}\rightarrow 0$, the term involving friction becomes dominant and we will retrieve Eq.~\ref{eq:thetaMaxDist}. 
Note that a similar result in the case without friction was obtained in Ref. \cite{AshwinPRE2013}, on the basis of an energy minimization argument.

\subsection{Macroscopic Failure: Slip-line Instability Analysis}\label{sec:MFSlipLineAnalysis}
In a highly simplified setting, a fully operating band may be idealized as a quasi-linear object made up of evenly distributed Eshelby events. 
Using an eigenmode based strategy, we now propose a simple theoretical argument that makes predictions on the band alignment. 
Point-like events with volume $a^d$ and amplitude $\epsilon^*$ are evenly positioned at points $\vec{r}_i$ on a line at an angle $\alpha$ with the $x$ axis: 
\begin{equation}
	\epsilon_{\alpha\beta}(\vec{r}) = \epsilon^*_{\alpha\beta}~a^d\sum_i\delta(\vec{r}-\vec{r}_i).
\end{equation}
Here $\delta(...)$ denotes the delta function.

Taking a Fourier transform, the $q$-space representation reads
\begin{equation}
	\label{eq:effSource}
	\hat\epsilon_{\alpha\beta}(\vec{q}) = Na^d \epsilon^*_{\alpha\beta}~\delta_{\vec{\hat{q}}.\vec{n}},
\end{equation}
where $\vec{n}=(\text{cos}\alpha,~\text{sin}\alpha)$ indicates the band direction, $\vec{\hat{q}}$ is the unit vector along $\vec{q}$, $\delta$ is the Kronecker symbol, and ${N}$ is the total  number of events \footnote{in order to keep the strain finite in the continuum limit, $Na^d=$constant with $N\rightarrow\infty$ and $a^d\rightarrow 0$}.
Here $\hat\epsilon_{\alpha\beta}(\vec{q})$ takes a localized form as well but at a direction perpendicular to the band orientation in real space.
Assuming pure shear distortion for the transformations, \emph{i.e.} $\epsilon^*_{\alpha\beta}=\epsilon^*(\delta_{\alpha x}\delta_{\beta x}-\delta_{\alpha y}\delta_{\beta y})$, it follows that
\begin{eqnarray}
	\delta p(\vec{q}) &=& -Na^d\frac{2\mu\epsilon^*}{1+\frac{\mu}{K}}~\text{cos}~2\theta ~\delta_{\vec{\hat q}.\vec{n}},  \\ 
	\delta\sigma(\vec{q}) &=& +Na^d[2\mu\epsilon^*-\frac{2\mu\epsilon^*}{1+\frac{\mu}{K}}~\frac{1}{2}(1+\text{cos}~4\theta)]~\delta_{\vec{\hat q}.\vec{n}}, \nonumber
\end{eqnarray} 
with $\theta$ the wave vector angle.
The above equations follow from operating the Oseen tensor on the effective source field defined by the line of shear transformations (see Appendix \ref{sec:stz}).

It is clear that the solutions in Fourier space will only depend on the direction $\theta$ (not on $|\vec{q}|$) and take the exact same form as the effective source in Eq.~\ref{eq:effSource}, up to a normalization factor. 
In other words, the slip line does not induce any stress redistribution outside of the line itself, and the  real-space response function is  localized along the slip line at angle $\alpha$
\begin{eqnarray}
	\label{eq:pressSigma}
	\delta p(\vec{r}) &=& \frac{2\mu\epsilon^*}{1+\frac{\mu}{K}}~\text{cos}~2\alpha~a^d\sum_i\delta(\vec{r}-\vec{r}_i)\\
	\delta\sigma(\vec{r}) &=& [2\mu\epsilon^*-\frac{2\mu\epsilon^*}{1+\frac{\mu}{K}}\frac{1}{2}(1+\text{cos}~4\alpha)]~a^d\sum_i\delta(\vec{r}-\vec{r}_i). \nonumber
\end{eqnarray}

As a result, the change in the yield function $\delta f_y$  is zero everywhere except on the slip line. We now argue that the most unstable lines will be those that maximize 
 $\delta f_y$, which is  positive inside the band. This corresponds to a maximum amplification of the local ``damage'' that is caused by the operating shear band for an infinitesimal strain. The shear band angle $\theta_\text{sh}$ is thus obtained by maximizing $\delta f_y$  with respect to $\alpha$:
\begin{equation}
	\label{eq:thetaSH}
	\frac{\partial}{\partial \alpha}\delta f_y|_{\alpha=\theta_\text{sh}}=0: \text{cos}~2\theta_\text{sh} = -\frac{1}{2}\text{sin}~\phi.
\end{equation}
This angle lies between $45^{\circ}\le\theta_\text{sh}\le60^{\circ}$, and for small $\phi$ is approximately $\theta_\text{sh}\approx 45^{\circ}+\frac{\phi}{4}$.
The combination of the pure shear deformation and volumetric change in the transformation zones gives 
\begin{equation}
	\label{eq:thetaSHDilt}
	\text{cos}~2\theta_\text{sh} = -\frac{1}{2}(\frac{\epsilon_v^*}{2\epsilon^*}+\text{sin}~\phi).
\end{equation}

Inserting Eq.~\ref{eq:thetaSH} into Eq.~\ref{eq:pressSigma} and integrating $\delta p$ over the active domain, it becomes evident that the total change in pressure is negative.
This will effectively reduce the shear strength  given by the Mohr-Coulomb yield surface inside the localization zone.
With such a mechanism,  yielded zones become likely places where next events tend to localize, hence making possible a permanent band-like formation.
This is compatible with the  numerical observations presented below  in Sec.~\ref{sec:bandStruct}.

From our discussions above it follows that, theoretically, $\theta_\text{max}$, the direction resulting from expected elastic-type correlations, will differ from the slip-line angle $\theta_\text{sh}$. 
This discrepancy appears to be consistent  with the observations made in the granular experiment of Le Bouil \emph{et al}. \cite{le2014emergence}. 
Note that the interpretation proposed by these authors was initially based on purely elastic considerations, with no allusion to the local friction. In this analysis, a large local dilatancy had to be assumed in order to make a sensible prediction of $\theta_\text{max}$. Our approach, by introducing the local friction angle, allows one to obtain a similar order of magnitude without invoking a large dilation.
In a follow-up work \cite{mcnamara2016eshelby}, it was suggested that the material anisotropy (or its combination with the volumetric strain) may explain the characteristic angle.  
The latter assumption was largely attributed to the presence of \emph{force chains} that build up upon loading granular solids \cite{karimi2011local}. 
Our interpretation is somewhat different in that it is entirely based upon the dominant role of the friction angle on correlation patterns, and in addition makes a specific prediction for the orientation of the permanent shear bands.
\section{Simulation Details}\label{sec:numMod}

In order to test the predictions of the theoretical discussion above, we have made use of theb Finite Elements based version of elasto-plastic models  that we originally established in \cite{KarimiPRE2016, karimi2017inertia}. In our previous work, this model was used to study the deformation of amorphous media in which the local failure is governed by a standard,
maximum shear stress criterion. Here, in order to follow the hypothesis of the previous section, the microscopic failure criterion for each element will be a Mohr-Coulomb condition. 
Each material point is therefore assigned with the shear strength parameters $c$ and $\phi$; while the former is randomly chosen from an exponential distribution with the mean value $\bar c$, we allocate no disorder to the latter.

\begin{figure}
	\includegraphics[width=8.6cm]{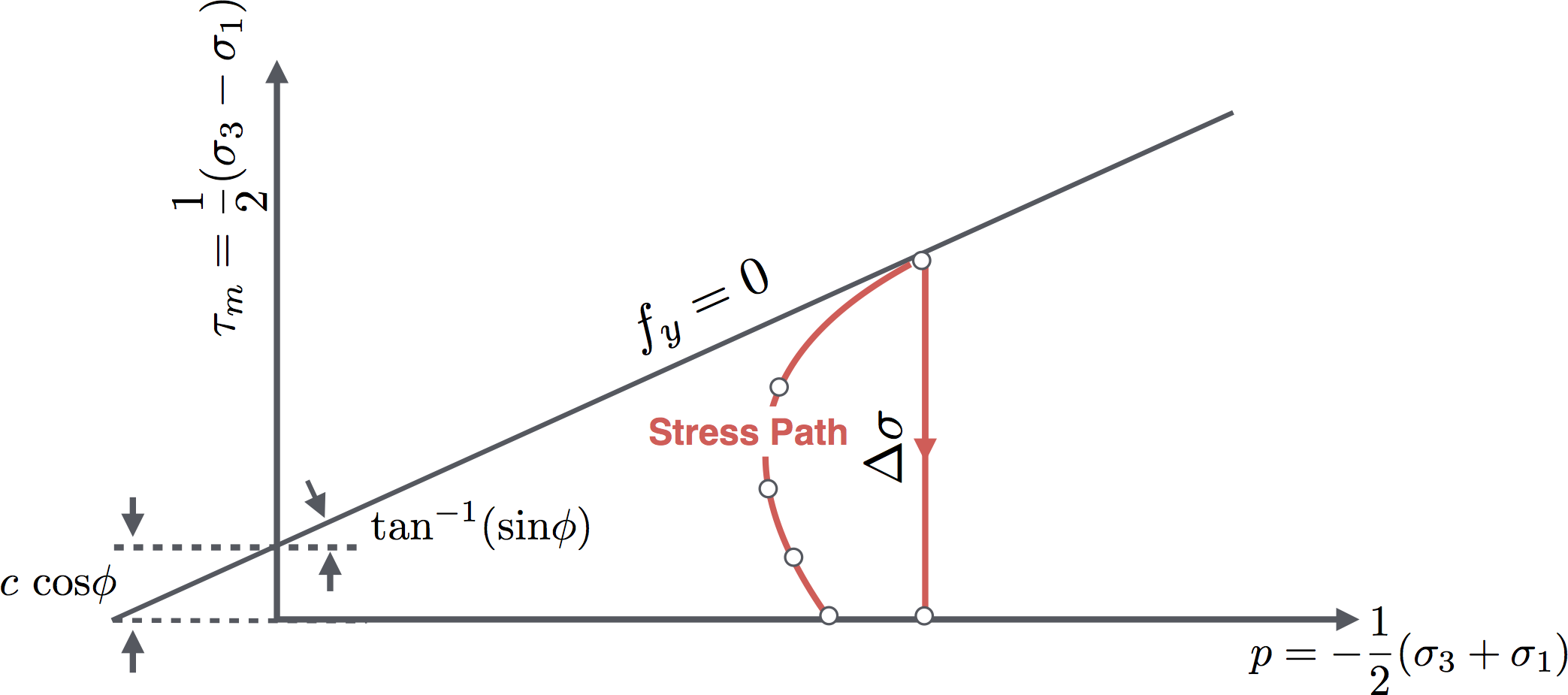}
	\caption{The Mohr-Coulomb failure envelope. The red curve and white dots sketch the stress trajectory and state of stress at a material point of interest.} 
	\label{fig:mohrColumb1}
\end{figure}

Below the failure limit at the material point in question, set by $f_y=0$ in the pressure-shear stress plane, the \emph{local} stress trajectory is regulated by the imposed drift and non-local elastic interactions.
We further presume a linear isotropic elastic response for the pre-failure dynamics. 
Figure \ref{fig:mohrColumb1} plots $\tau_m=\frac{1}{2}(\sigma_3-\sigma_1)$ against $p=-\frac{1}{2}(\sigma_3+\sigma_1)$ and represents any stress state by a stress point.
Here $\sigma_1$ and $\sigma_3$ denote major and minor principle stresses, respectively.
Upon yielding, the stress point will take on a path which is perpendicular to the $p$-axis and relax visco-elastically \cite{karimi2017inertia} toward a point representing the final state of stress. 
The released shear stress $\Delta\sigma$ will create a localized net force that perturbs the force equilibrium in the medium. 


Simulations  of shear deformations were performed by applying an area preserving \emph{axial} shear rate $\dot\epsilon \sim 10^{-4}$ to an $L\times L$ periodic cell.
An irregular set of triangular elements with average size $h=\frac{L}{80}$ was used to discretize the domain.
The rate unit (inverse timescale) is set by the shear wave velocity $c_s=(\mu/\rho)^\frac{1}{2}$ divided by $L$ with $\rho$ the mass density.
The rate of strain chosen is slow enough to ensure \emph{quasi-static} conditions, \emph{i.e.} the results are insensitive to a reduction in the value of the rate.
We also set $K/\mu=2$ in the elastic regime, corresponding to a Poisson ratio of $\nu\approx0.33$.

By setting a high value of the damping rate (in comparison with the vibrational frequency), we moreover impose an overdamped behavior during the relaxation phases, thus ensuring that inertial effects are negligible \cite{karimi2017inertia}.
Prior to shearing, samples were prepared with random stresses assigned to each block followed by an equilibration within a purely elastic framework (no plastic events allowed) that resulted in a state of mechanical equilibrium. 
%
\section{Numerical Results} \label{sec:NumRes}
\begin{figure}
	\begin{center}
		\begin{overpic}[width=8.6cm]{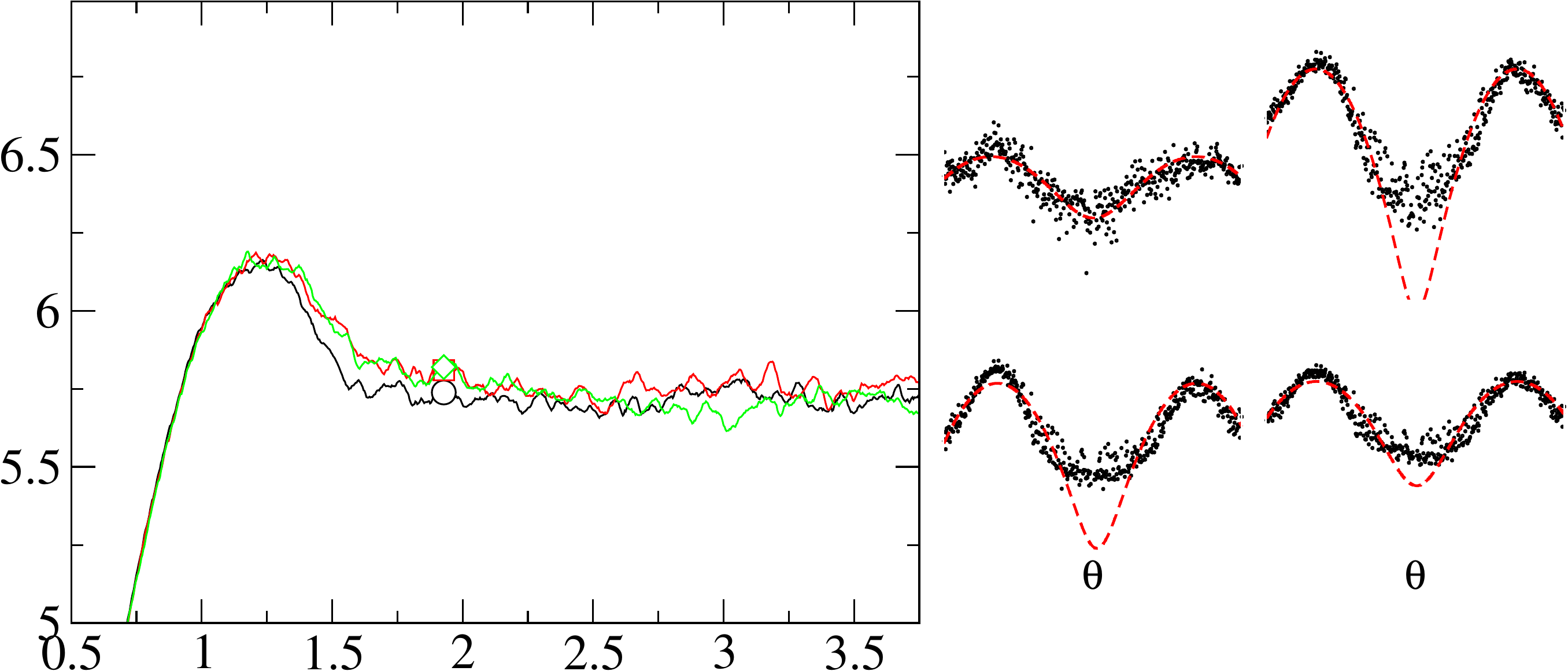}
			\put (32,-3) {\small$\epsilon$\tiny{/}\normalsize$\epsilon_y$} 
			\put (-3,21) {\small\begin{turn}{90}$\sigma$\tiny{/}\normalsize$\bar c$\end{turn}}
 		    \put(60.19,23.7){\includegraphics[width=1.61cm,frame]{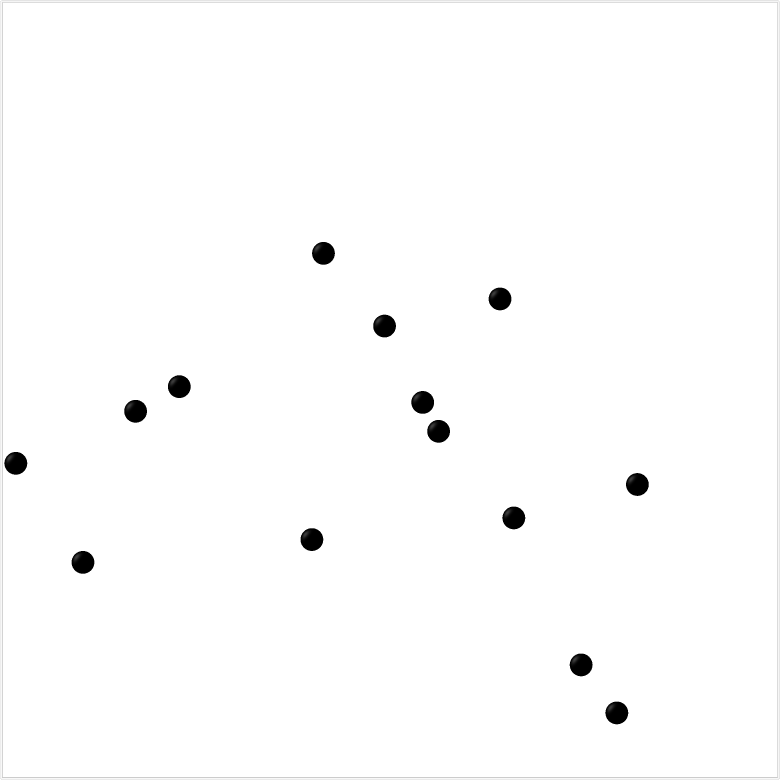}} 
			 \put(80.8,23.7){\includegraphics[width=1.61cm,frame]{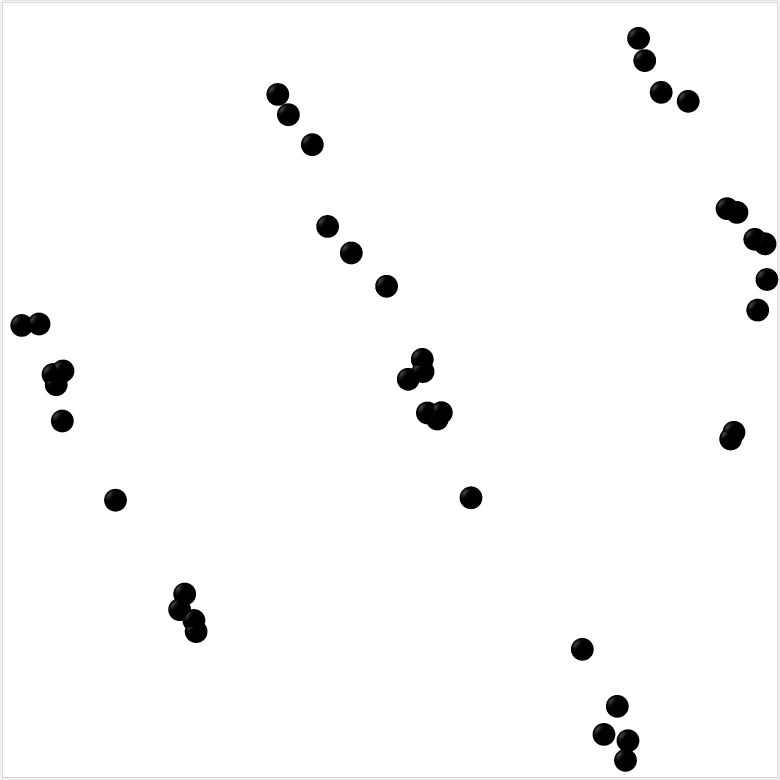}} 
			 \put(60.19,3.25){\includegraphics[width=1.61cm,frame]{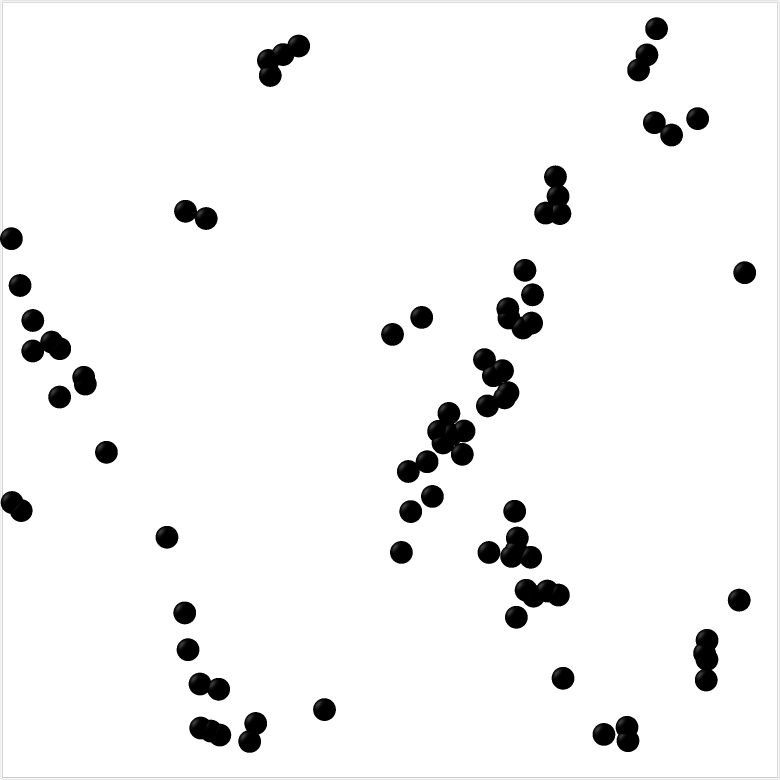}}
			 \put(80.8,3.25){\includegraphics[width=1.61cm,frame]{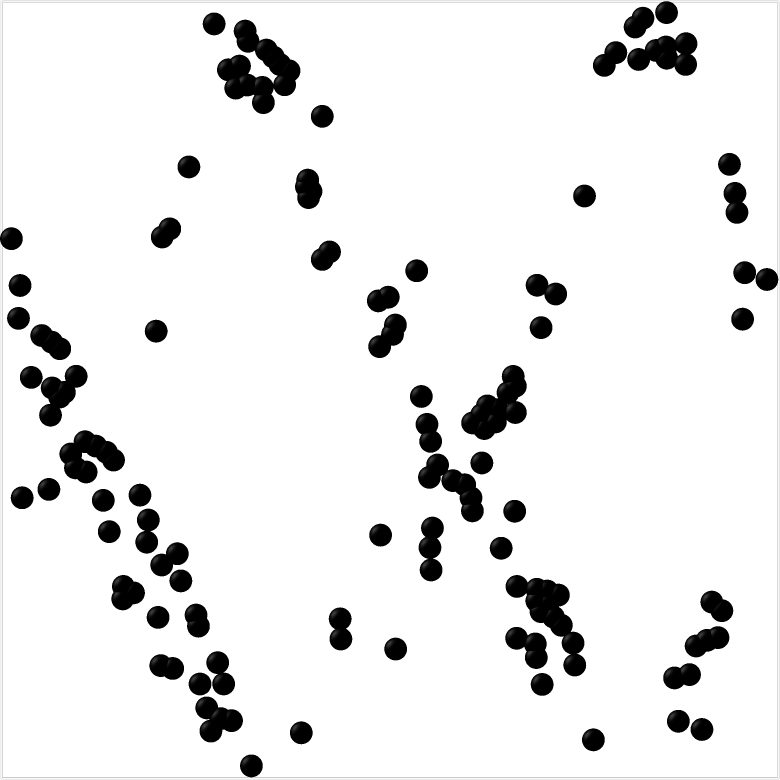}}
			\put (4.5,39.7) {\sffamily\setlength{\fboxsep}{0pt}\colorbox{black}{\strut\bfseries\textcolor{white}{\small$(a)$}}} 
			\put (74.5,39.6) {\sffamily\setlength{\fboxsep}{0pt}\colorbox{black}{\strut\bfseries\textcolor{white}{\small$(b)$}}} 
			\put (95.2,39.6) {\sffamily\setlength{\fboxsep}{0pt}\colorbox{black}{\strut\bfseries\textcolor{white}{\small$(c)$}}} 
			\put (74.3,19) {\sffamily\setlength{\fboxsep}{0pt}\colorbox{black}{\strut\bfseries\textcolor{white}{\small$(d)$}}} 
			\put (95.2,19) {\sffamily\setlength{\fboxsep}{0pt}\colorbox{black}{\strut\bfseries\textcolor{white}{\small$(e)$}}} 
			\put(40,25){\includegraphics[width=1.4cm]{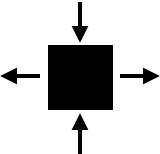}} 
			\put (47.5,32) {\color{white}{\small$\dot\epsilon$}} 
		\end{overpic}
	\end{center}
	\caption{Results of bi-axial shearing tests. (a) Stress-strain curves for three samples (denoted by different symbols) tested under the same value of pressure $p=8\bar c$ and $\phi=65^\circ$. The shear stress $\sigma$ is measured in units of $\bar c$. (b-e) Spatial maps of active sites illustrated by solid dots at $\epsilon$\tiny{/}\normalsize$\epsilon_y=1.25, 1.6, 2.5,$ and $3.7$. Each map corresponds to the full system size $L$.}
	\label{fig:stressStrain}
\end{figure}
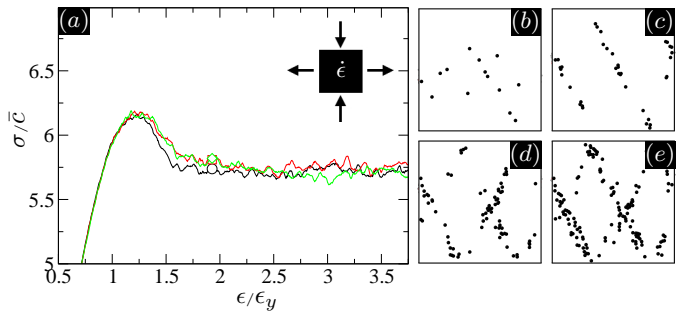

The results of shearing tests can be used to determine the \emph{bulk} shear strength characteristics together with the structure of interactions between transient slip events.
A number of tests has been performed at the same (initial) pressure $p=8\bar c$ and $\phi=65^\circ$, and the resulting average stress strain curve  is displayed in Fig.~\ref{fig:stressStrain}(a).
For the given $p$ and $\phi$, the material shear strength has the average value of $\sigma_y\approx 8\bar c$.
We also report $\epsilon$ in units of the yield strain $\epsilon_y=\frac{\sigma_y}{2\mu}$. 
A peak stress is typically reached around $\epsilon\approx\epsilon_y$ in every sample followed by a reduction in strength as the loading continues.
With increasing strain, the shear strength ultimately falls to a residual value at large deformations.


A strong plastic activity in the form of extended linear structures is present at all times after the initial yielding for $\epsilon>\epsilon_y$.
Spatial maps of active sites in Fig.~\ref{fig:stressStrain}(b-e) demonstrate the highly intermittent and \emph{non}-local nature of bursts during plastic flow (see also the movie in the Supplementary Materials \footnote{Follow \href{https://drive.google.com/open?id=0BySY_gg-2V8pZEFhT3RQQmJzU0k}{the link} to play the movie for the evolution of activity maps as in Figs.~\ref{fig:stressStrain}($b-e$) at different loading steps (the imposed strain is also indicated). The color map is based on the height of each active site. The data was visualized by OVITO \cite{stukowski2009visualization}}).
Following the stress peak, multiple shearing bands start to evolve within which most of plastic activities are taking place. 
The orientation of these bands are quite distinct from maximum shearing directions, $45^\circ$ or $135^\circ$ in our loading set-up.
The quantification of the slip directions will be the object of Sec.~\ref{sec:bandStruct}, which presents an analysis similar in spirit to \cite{nguyen2016experimental}.
 
%
%

\subsection{Structural Characterization}\label{sec:bandStruct}
Here we focus on the number density of active sites and associated spatial correlations so as to quantify the spatial structure of  the plastic activity.   
Let $\rho(\vec{r})=\displaystyle \sum_i \delta(\vec{r}-\vec{r}_i)$ where $\vec{r}_i$ is the position vector of a plastically active  zone with index $i$ and $\delta(...)$ denotes a delta function. Integrating $\rho(\vec{r})$ over the entire volume $V$ provides the total number of plastic sites $N$ at the current time or strain.
We are interested in the variations of $\rho(\vec{r})$ about its average value $\rho=\frac{N}{V}$. These variations can be characterized by a 
 two-point density correlation function  between two different positions  $\vec{r}$ and $\vec{r}^\prime$ in space:
$S(\vec{r},\vec{r}^\prime)\doteq\langle [\rho(\vec{r})-\rho] [\rho(\vec{r}^\prime)-\rho]\rangle$. 
Here the angular brackets $\langle ... \rangle$ correspond to an average over different realizations and a spatial average. As a result of translational invariance
$S(\vec{r},\vec{r}^\prime) \equiv S(\vec{r}-\vec{r}^\prime)$.
A Voronoi cell analysis was performed that enabled interpolations of the density field onto fine regular grids. 
We subsequently used a 2d Fourier transform in order to compute the  correlations.

\begin{figure}
	\begin{center}
		\begin{overpic}[width=0.155\textwidth,frame]{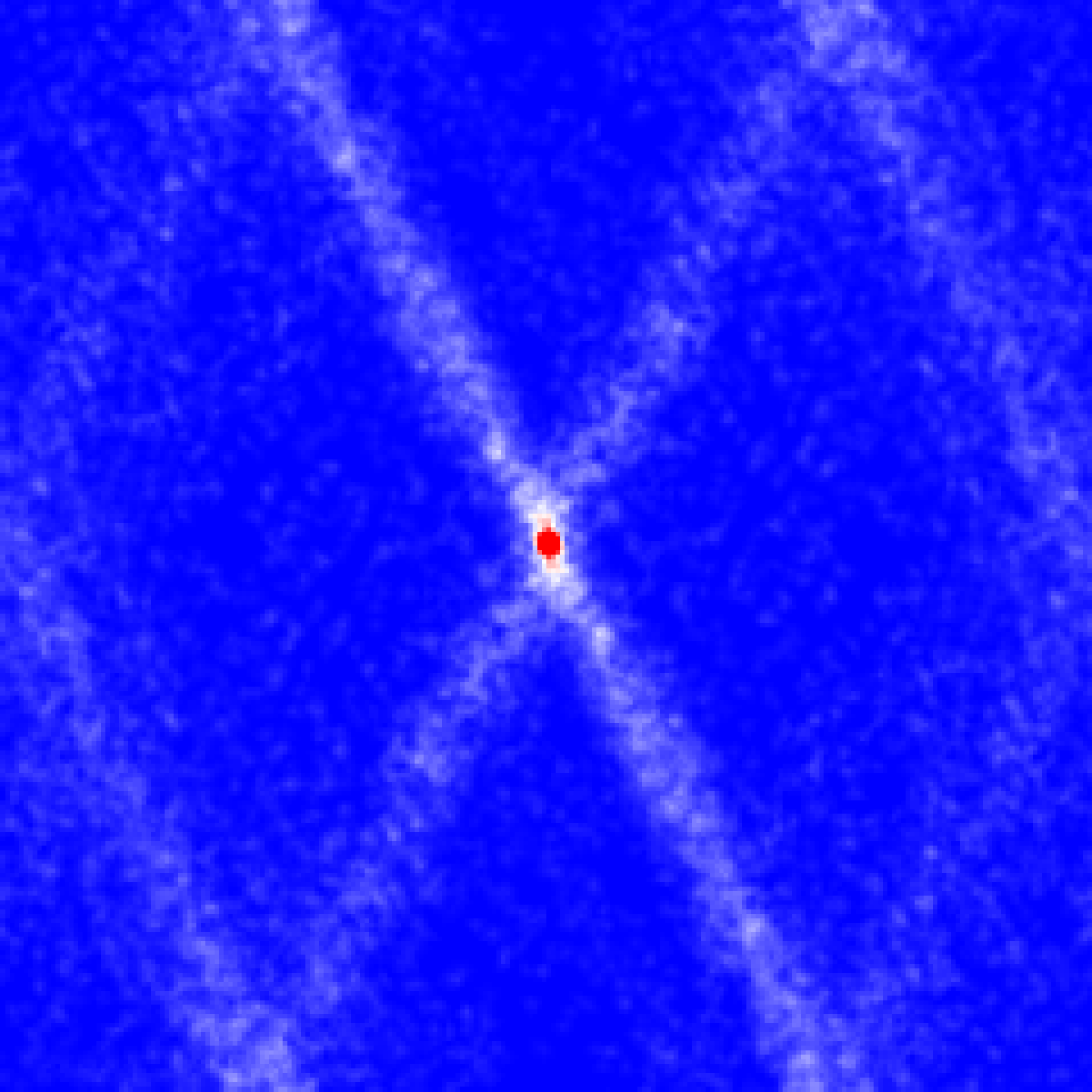}
			\put (.5,90.5) {\sffamily\setlength{\fboxsep}{0pt}\colorbox{black}{\strut\bfseries\textcolor{white}{\small$(a)$}}} 
		\end{overpic}
		\begin{overpic}[width=0.155\textwidth,frame]{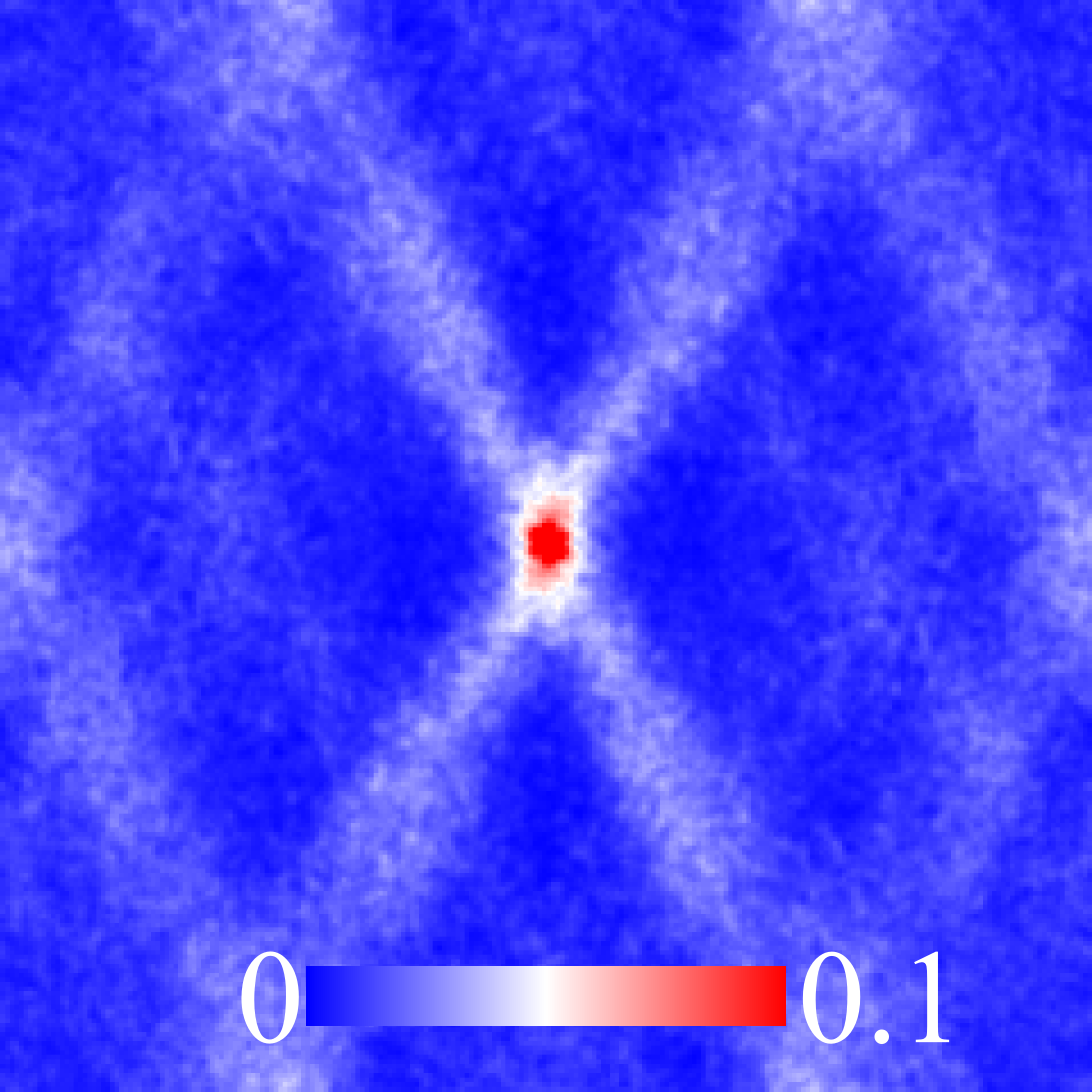}
			\put (0.5,90.5) {\sffamily\setlength{\fboxsep}{0pt}\colorbox{black}{\strut\bfseries\textcolor{white}{\small$(b)$}}} 
                
		\end{overpic}
		\begin{overpic}[width=0.155\textwidth,frame]{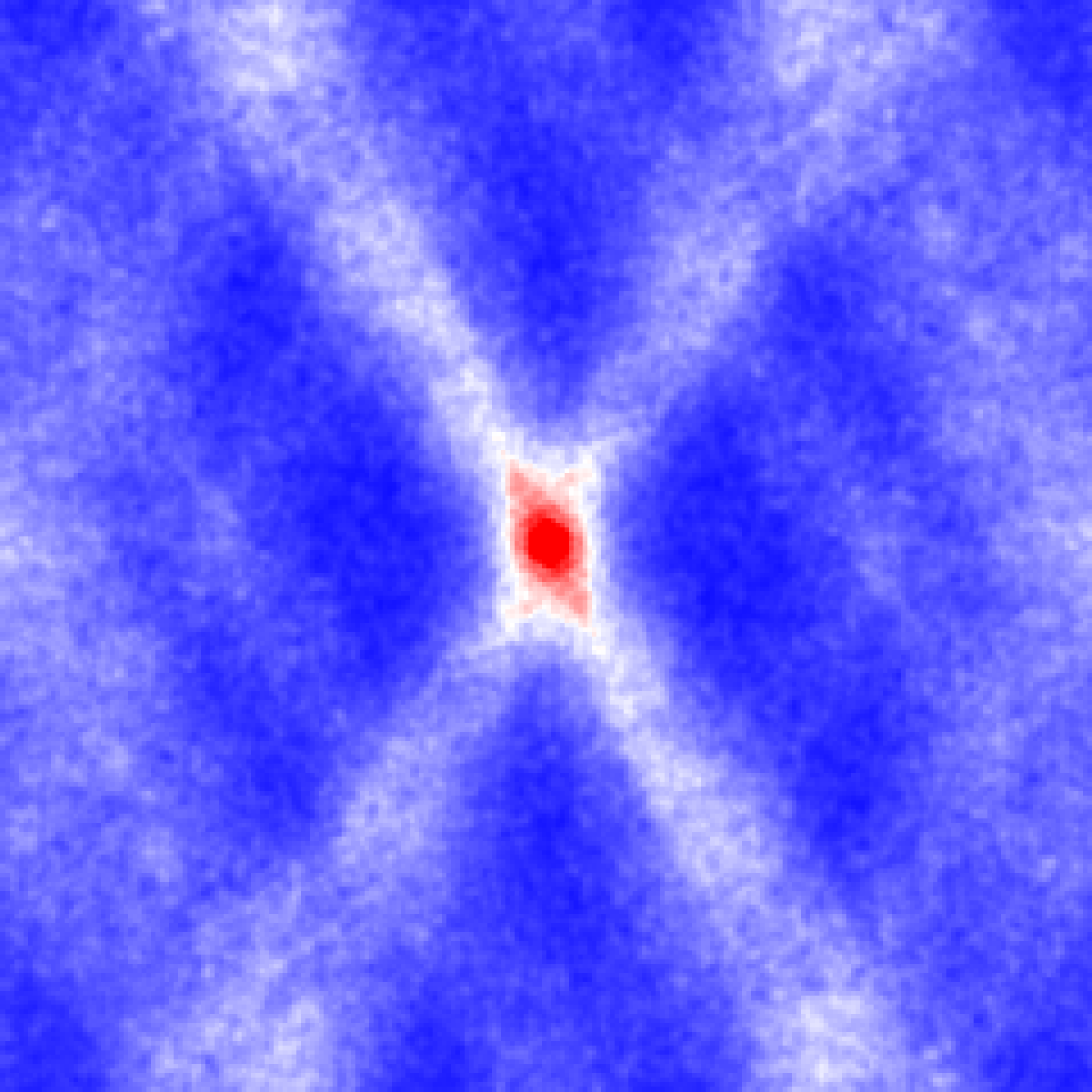}
			\put (0.5,90.5) {\sffamily\setlength{\fboxsep}{0pt}\colorbox{black}{\strut\bfseries\textcolor{white}{\small$(c)$}}} 
			\begin{tikzpicture} 
				\coordinate (a) at (0,0); 
				\node[] at (a) {\tiny.};
				\coordinate (center) at (1.26,1.28); 
				\draw[dashdotted] (center) -- ( $ (center) + 1.11*(1,1.259) $); 
				\draw[dashdotted] (center) -- ( $ (center) + (0.5,0) $); 
				\draw ($(center)+(.2,0)$) arc(0:51.548:2mm); 
				\draw[fill=black] ($(center)+(.34,.05)$) rectangle ($(center)+(.34,.05)+(0.47,0.3)$); 
				\node[text=white] at ($(center)+(.6,.21)$) {\small{$52^\circ$}}; 
				\draw (center) -- ( $ (center) + .86*(-1,1.630) $); 
				\draw (center) -- ( $ (center) + (-0.5,0) $); 
				\draw
				 ($(center)+(-0.1,0.16)$) arc(121.527:180.0:2mm); 
				\draw[fill=black] ($(center)+(.34-1.1,.05)$) rectangle ($(center)+(.34-1.1,.05)+(0.47,0.3)$); 
				\node[text=white] at ($(center)+(.6-1.1,.21)$) {\small{$59^\circ$}}; 
			\end{tikzpicture}

		\end{overpic}
	\caption{Density correlation function $S(\vec{r}-\vec{r}^\prime)$ at multiple strains (a) $\epsilon$\tiny{/}\normalsize$\epsilon_y=1.4$ (b) $\epsilon$\tiny{/}\normalsize$\epsilon_y=2.5$ (c) $\epsilon$\tiny{/}\normalsize$\epsilon_y=3.7$. Here $p=8\bar c$ and $\phi=65^\circ$. The scale of each density map is $L$. The solid and dashed-dotted lines indicate the theoretical predictions for $\theta_\text{sh}$ and $\theta_\text{max}$, respectively.}
	\label{fig:structFact}
	\end{center}
\end{figure}
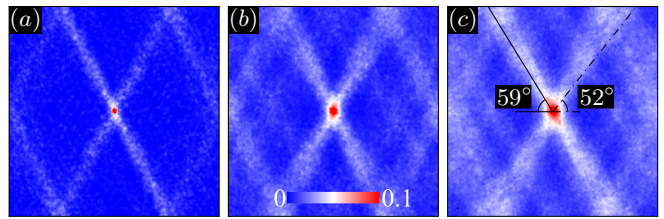

We shall naturally expect that density fluctuations are highly correlated along shear band directions, inducing strong anisotropies in $S(\vec{r}-\vec{r}^\prime)$. 
Figure \ref{fig:structFact} displays the evolution of density correlations at different loading stages averaged over 16 samples at $p=8\bar c$ and $\phi=65^\circ$.  
Angular symmetries seen in the post-failure regime are fairly stable features showing only weak fluctuations with increasing strain.
The banded regions contain correlations that are explicitly longer-ranged (as opposed to other orientations) indicating the system-spanning nature of the localization. 
%
%

\begin{figure}
	\begin{center}
		\begin{overpic}[width=8.6cm]{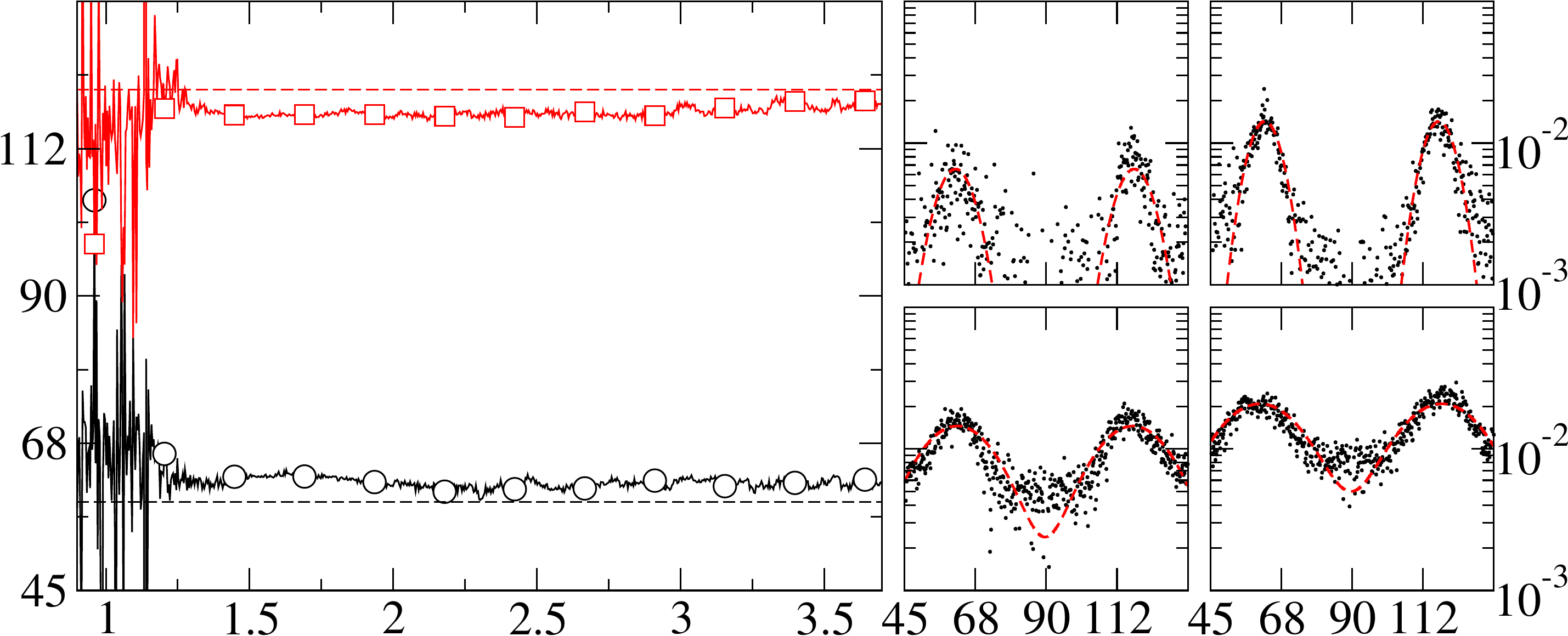}
			\put (32,-3) {\small$\epsilon$\tiny{/}\normalsize$\epsilon_y$} 
			\put (-3,21) {\small\begin{turn}{90}$\theta_\text{sh}$\end{turn}}
			\put (67,-3) {\small$\theta$} 
			\put (86,-3) {\small$\theta$} 
			\put (100,8.9) {\small\begin{turn}{90}$S_{(\theta)}$\end{turn}} 
			\put (100,28) {\small\begin{turn}{90}$S_{(\theta)}$\end{turn}} 
			\put (51.2,37.5) {\sffamily\setlength{\fboxsep}{0pt}\colorbox{black}{\strut\bfseries\textcolor{white}{\small$(a)$}}} 
%
			\put (71.3,37.5) {\sffamily\setlength{\fboxsep}{0pt}\colorbox{black}{\strut\bfseries\textcolor{white}{\small$(b)$}}} 
			\put (90.8,37.5) {\sffamily\setlength{\fboxsep}{0pt}\colorbox{black}{\strut\bfseries\textcolor{white}{\small$(c)$}}} 
			\put (70.9,17.9) {\sffamily\setlength{\fboxsep}{0pt}\colorbox{black}{\strut\bfseries\textcolor{white}{\small$(d)$}}} 
			\put (90.7,17.9) {\sffamily\setlength{\fboxsep}{0pt}\colorbox{black}{\strut\bfseries\textcolor{white}{\small$(e)$}}} 
		\end{overpic}
	\end{center}
	\caption{Anisotropic part of the density correlation function $S(\theta)$ at $p=8\bar c$ and $\phi=65^\circ$. ($a$) Locations of the peaks in $S(\theta)$ denoted by $\theta_\text{sh}$ versus $\epsilon$\tiny{/}\normalsize$\epsilon_y$. ($b-e$) $S(\theta)$ plotted against $\theta$ at $\epsilon$\tiny{/}\normalsize$\epsilon_y=1.4, 1.6, 2.5,$ and $3.7$. The dashed lines in the main plot designate theoretical predictions discussed in Sec.~\ref{sec:MFSlipLineAnalysis}. Gaussian fits in the insets are denoted by the dashed curves.}
	\label{fig:thetaStrain}
\end{figure}
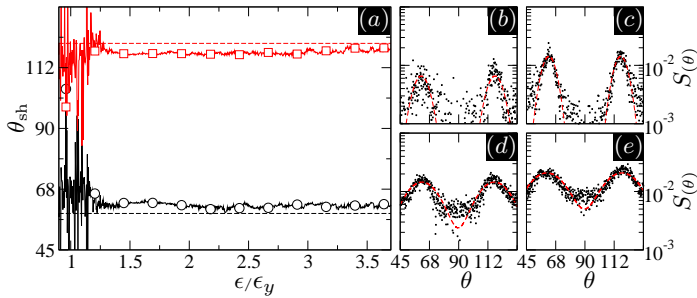

Figure~\ref{fig:thetaStrain} displays the anisotropic part $S(\theta)$, an averaged $S(\vec{r}-\vec{r}^\prime)$ over different distances $|\vec{r}-\vec{r}^\prime|$.
There are two marked maxima in every data set at $45^\circ<\theta<135^\circ$ that delineate the degree of anisotropy.
We quantify the positions of these peaks by fitting the data to a sum of two Gaussian peaks, as in Fig.~\ref{fig:thetaStrain}(b-e).
These fits enable us to obtain the locations of the peaks --denoted by $\theta_\text{sh}$-- and to follow their evolution upon shear loading.
This is shown in Fig.~\ref{fig:thetaStrain}(a).
Apart from the initial transient part prior to failure, the peak locations are essentially strain independent at larger strains.
The dashed lines in Fig.~\ref{fig:thetaStrain}(a) mark the directions of failure given by Eq.~\ref{eq:thetaMaxDist}.

The basis for the permanent localization observed here can be understood in simple terms.
An incident avalanche tends to \emph{depressurize} currently-damaged blocks, with a pressure drop proportional to $\text{sin}~\phi$, making them vulnerable spots against further deformations.
In the standard framework of elastoplastic models, it is possible to observe a similar behavior (however with the classical $45^\circ$ orientation for the shear band by adding some permanent or transient weakening mechanism to the model).
Examples include models based on the damage factor \cite{amitrano1999diffuse}, weakened stress thresholds \cite{vandembroucq2011mechanical}, or lingered restoration time \cite{martens2012spontaneous}, just to name a few.
In our simulations, the weakening is intrinsically contained in the pressure sensitivity of the failure criterion, and depends on the local friction angle.
We note that a relatively large friction angle $\phi$ was required to observe a permanent localization.

\subsection{Mohr-Coulomb Failure Envelope}\label{sec:MCFailureEnvelope}
\begin{figure}
	\begin{center}
		\begin{overpic}[width=8.6cm]{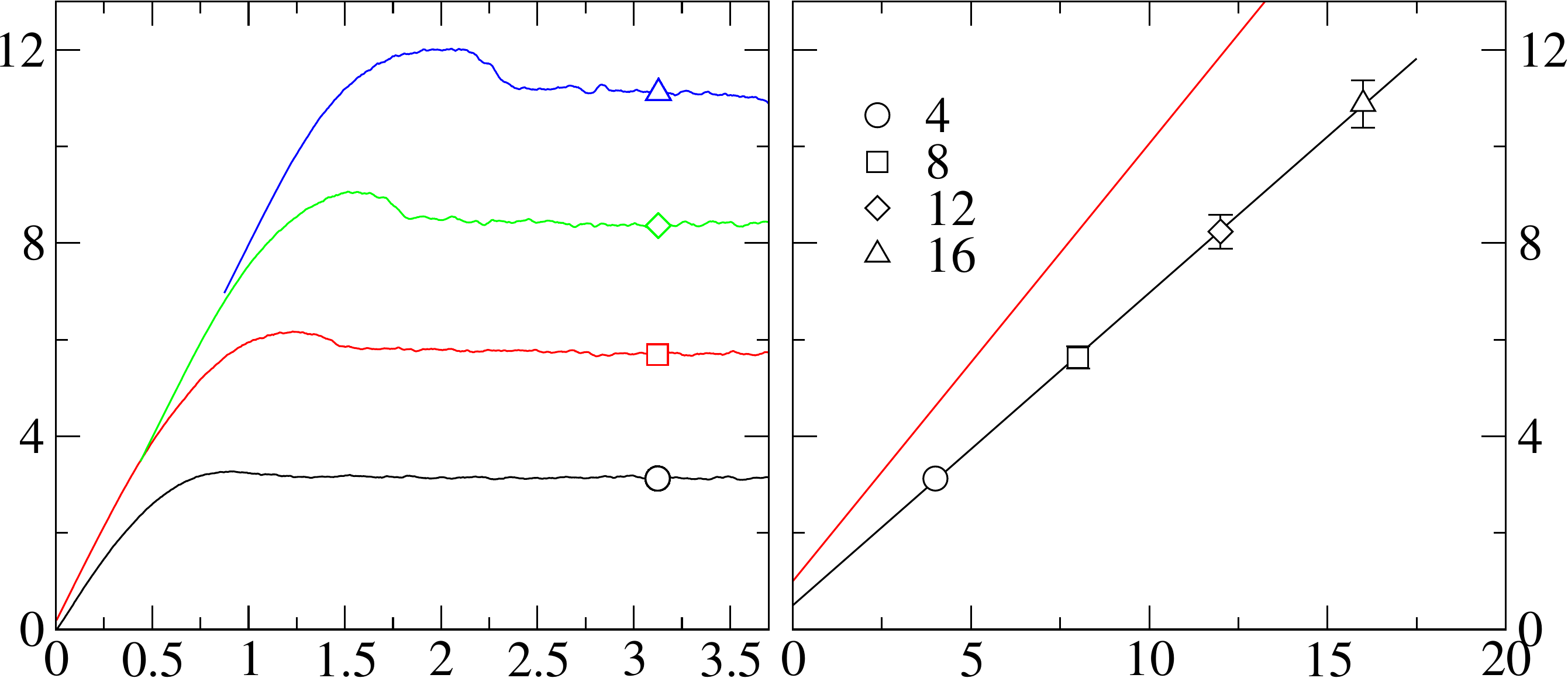}
			\put (26,-3) {\small$\epsilon$\tiny{/}\normalsize$\epsilon_y$} 
			\put (-3,21) {\small\begin{turn}{90}$\sigma$\tiny{/}\normalsize$\bar c$\end{turn}}
			\put (44,40.1) {\sffamily\setlength{\fboxsep}{0pt}\colorbox{black}{\strut\bfseries\textcolor{white}{\small$(a)$}}} 
			\put (71,-3) {\small$p$\tiny{/}\normalsize$\bar c$} 
			\put (100,21) {\small\begin{turn}{90}$\sigma_r$\tiny{/}\normalsize$\bar c$\end{turn}} 
			\put (91.4,40.1) {\sffamily\setlength{\fboxsep}{0pt}\colorbox{black}{\strut\bfseries\textcolor{white}{\small$(b)$}}} 
			\put (56,39) {\small$p$\tiny{/}\normalsize$\bar c$} 
			\begin{tikzpicture} 
				\coordinate (ref) at (0,0); 
				\node[] at (ref) {\tiny.};
				\coordinate (a) at (5.95,1.81); 
				\coordinate (b) at ($ (a) + (0.25,0) $);
				\coordinate (c) at ($ (b) + (0,0.22) $); 
				\draw[line width=0.1mm] (a) -- (b); 
				\draw[line width=0.1mm] (b) -- (c); 
				\node at ($(b)-(0.1,0.1)$) {\tiny{1}}; 
				\node at ($(b)+(0.4,0.15)$) {\tiny{$\text{sin}40^{\circ}$}}; 
				\coordinate (ref) at (0,0); 
				\node[] at (ref) {\tiny.};
				\coordinate (a) at (5.95,2.56); 
				\coordinate (b) at ($ (a) + (0.25,0) $);
				\coordinate (c) at ($ (b) + (0,0.3) $); 
				\draw[line width=0.1mm,red] (a) -- (b); 
				\draw[line width=0.1mm,red] (b) -- (c); 
				\node at ($(b)-(0.1,0.1)$) {\color{red}\tiny{1}}; 
				\node at ($(b)+(0.4,0.15)$) {\color{red}\tiny{$\text{sin}65^{\circ}$}}; 
			\end{tikzpicture}
		\end{overpic}
	\end{center}
	\caption{Measurement of the bulk shear parameter $\phi_r$. (a) stress-strain curves for multiple values of the confining pressure $p$ and $\phi=65^\circ$. The symbols lay out stress points used to draw the MC failure envelope. (b) stress points commensurate with the residual strengths $\sigma_r$ plotted on the shear-pressure plane. The lines are MC linear fits with slopes $\text{sin}~\phi_r$ and $\text{sin}~\phi$ .} 
	\label{fig:FailureEnvelope}
\end{figure}
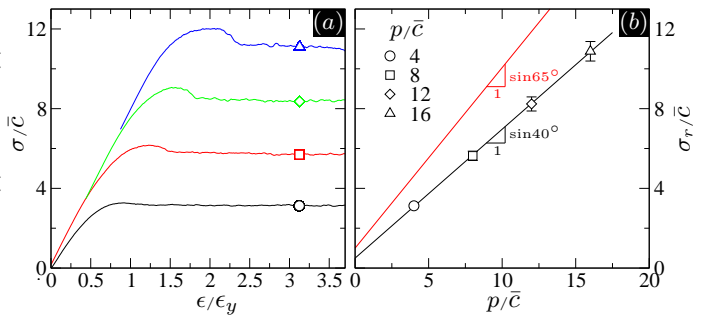

Testing several samples each under different confining pressures enables the determination of a \emph{macroscopic} failure envelope and the \emph{bulk} shear strength parameters. Figure~\ref{fig:FailureEnvelope} illustrates the results in a series of tests performed at $p$\tiny{/}\normalsize$\bar c=4, 8, 12,$ and $16$ and $\phi=65^\circ$.
Every set of tests was carried out on 16 independent samples.
The data points in Fig.~\ref{fig:FailureEnvelope}(b) represent the states of stress on the shear-pressure plane.
These stress points correspond to the ultimate strength \footnote{Very similar results would be obtained by using the peak values in the stress strain curves in Fig.~\ref{fig:FailureEnvelope}(a)}, marked by the symbols in Fig.~\ref{fig:FailureEnvelope}(a). 
Assuming a \emph{global} Mohr-Coulomb criterion, the residual strengths are expressed by 
%
\begin{equation}
\sigma_r=p~\text{sin}~\phi_r+c_r~\text{cos}~\phi_r,
\end{equation}
where $\phi_r$ denotes the value of the bulk friction angle and $c_r$ is the macroscopic cohesion.

The parameters can be determined empirically by making linear fits to the points. 
The values of the measured parameters are:
\begin{equation}
\nonumber
\phi_r=40^\circ;~c_r=0.5~\bar c.
\end{equation} 
As stipulated by the Mohr-Coulomb phenomenology, the theoretical angle between the major principal stress direction and the plane of failure must be $\theta_\text{MC}=45^{\circ}+\frac{\phi_r}{2}$.
Inserting $\phi_r=40^\circ$, it follows that $\theta_\text{MC}=65^\circ$ which is off from $\theta_\text{sh}$ by about $6^\circ$. 
Therefore, we find that applying the macroscopic Mohr-Coulomb theory using the observed values of the macroscopic failure envelope tends to overestimate the actual inclination of the  yield surface, in agreement with experimental observations \cite{bardet1990comprehensive}.
%
\subsection{Pre-failure  Patterns of Plastic Activity}
Beyond the  yield strain, correlations in plastic activity are easily identified based on the analysis of a snapshot taken at a single time or strain, as described above using the function $S(\vec{r}-\vec{r}^\prime)$. This is related to the existence of well correlated linear regions of plastic activity that are operating simultaneously, eventually giving rise to localized shear bands. In the pre-failure regime, however, plastic activity is much more scattered, and the study of a single configuration does not reveal any established pattern. A calculation of $S(\vec{r}-\vec{r}^\prime)$ does not allow one to identify any preferred direction.  The structure of the plastic activity, however, can be revealed by the study of two time correlations between configurations separated by a fixed strain interval $\Delta \epsilon$. In practice, the statistics is improved by averaging over different simulation and performing an average over a strain window for the initial configuration. We define 
\begin{equation}
C(\vec{r}-\vec{r}^\prime,\Delta \epsilon) \doteq\langle \frac{1}{\epsilon_2-\epsilon_1} \int_{\epsilon_1}^{\epsilon_2} d\epsilon\rho(\vec{r},\epsilon+\Delta\epsilon)\rho(\vec{r}^\prime,\epsilon)\rangle.
\end{equation}
Here the brackets denote the average over different realizations. The averaging interval $[\epsilon_1,\epsilon_2]$ is taken before the strain peak, $\epsilon_1=0.25\epsilon_y$ and $\epsilon_2=0.7\epsilon_y$,  in order to avoid the contamination of the correlation functions by the formation of permanent shear bands in the post yield regime.

\begin{figure}
	\begin{center}
		\begin{overpic}[width=0.155\textwidth,frame]{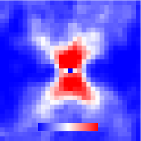} 
			\put (.5,90.5) {\sffamily\setlength{\fboxsep}{0pt}\colorbox{black}{\strut\bfseries\textcolor{white}{\small$(a)$}}} 
			\begin{tikzpicture} 
				\coordinate (a) at (0,0); 
				\node[] at (a) {\tiny.};
				\coordinate (center) at (1.26,1.28); 
				\draw[dashdotted] (center) -- ( $ (center) + 1.11*(1,1.259) $); 
				\draw[dashdotted] (center) -- ( $ (center) - 1.11*(1,1.259) $); 
				\draw[dashdotted] (center) -- ( $ (center) + (0.5,0) $); 
				\draw ($(center)+(.2,0)$) arc(0:51.548:2mm); 
				\draw[fill=black] ($(center)+(.34,.05)$) rectangle ($(center)+(.34,.05)+(0.47,0.3)$); 
				\node[text=white] at ($(center)+(.6,.21)$) {\small{$52^\circ$}}; 
				\draw[dashdotted] (center) -- ( $ (center) + 1.11*(-1,1.259) $); 
				\draw[dashdotted] (center) -- ( $ (center) - 1.11*(-1,1.259) $); 
				\node[text=white] at ($(center)+(-.85,-1.1)$) {\small $0$};
				\node[text=white] at ($(center)+(.85,-1.1)$) {\small $0.6$};
			\end{tikzpicture}
		\end{overpic}
		\begin{overpic}[width=2.83cm]{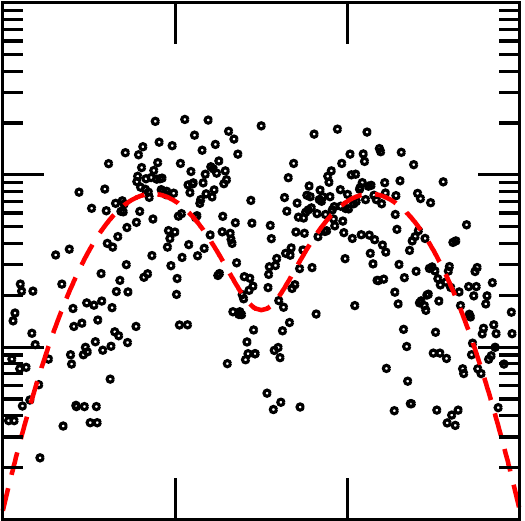}
			 \put(0,-10){\includegraphics[width=2.8cm]{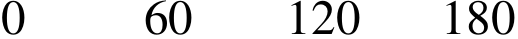}} 
			 \put(102,0){\includegraphics[height=3.0cm]{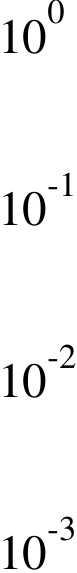}} 
			\put (0.5,90.0) {\sffamily\setlength{\fboxsep}{0pt}\colorbox{black}{\strut\bfseries\textcolor{white}{\small$(b)$}}} 
			\put (45,-16) {\small$\theta$} 
			\put (115,45) {\small\begin{turn}{90}$C_{(\theta)}$\end{turn}} 
		\end{overpic}
%
	\caption{(a) $C(\vec{r},\Delta\epsilon=5\times 10^{-4}\epsilon_y)$ at $p=8\bar c$ and $\phi=65^\circ$. The dashed-dotted lines indicate the theoretical prediction for $\theta_\text{max}$. Here the lengthscale is $\frac{L}{8}$. (b) Anisotropic portion $C(\theta)$ plotted against $\theta$. The Gaussian fit is represented by the dashed curve.}
	\label{fig:structFactTemp}
	\end{center}
\end{figure}
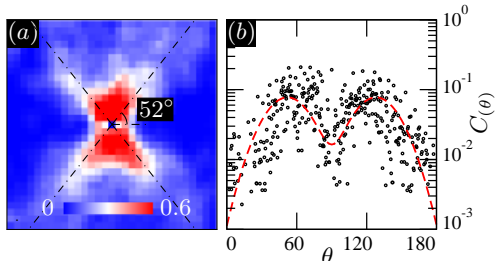

Statistically speaking, $C(\vec{r},\Delta \epsilon)$ will provide  spatial details about the most  likely position of an  event that is triggered   following a local slip event at the origin. In Fig.~\ref{fig:structFactTemp}(a), correlations are displayed for a small strain difference $\Delta\epsilon=5\times10^{-4}\epsilon_y$. The results exhibit a four-fold structure, which persists  up to a strain interval of the order $10^{-3}$ before fluctuations become uncorrelated at higher strain differences. 
The four-fold structure is of course expected from standard elasticity theory, and has been observed in experiments and simulations of a number of glassy systems. However, we observe here a marked deviation from the $45^\circ$ that would be predicted by pure elasticity, reflecting the influence of the friction angle in the yield criterion. 

In Fig.~\ref{fig:structFactTemp}(b), the angular dependence of $C(\vec{r},\Delta \epsilon)$ is shown together with a fit to a Gaussian function, similar to the one used for the analysis of $S(r, \theta)$. The fit function has a peak centered around $52^\circ$ which is in close agreement with the theoretical prediction $\theta_\text{max}$ 
Note that in order to reduce the noise in the data, the integration over $r$ was limited to small values of the distance $r<\frac{L}{16}$. 

The structure of correlations preceding the ultimate failure reflects the hypothesis of our model. 
A localized event induces stress fluctuations in the surroundings triggering plastic rearrangements nearby in the medium.
events are preferentially triggered in directions in which the changes to the yield function are maximal, as a result from non-local elastic couplings combined with a local frictional yielding rule.
While one may have expected that these fluctuations would strongly influence the collective behavior that emerges upon macroscopic failure, the analysis of the shear band orientations shows that they represent a distinct phenomenon, and must be analyzed within a more collective perspective.
%
%
\section{Conclusion} \label{sec:cnlsn}
In this work, we have taken a coarse-grained, mesoscopic view of the deformation in a disordered granular medium, by integrating the notion of a deformation taking place through local shear transformations with that of a local criterion of failure described by the Mohr-Coulomb friction condition. 
As a result, the mechanical response is described by three ingredients, the elastic moduli of the medium which, according to the Eshelby's picture, describe the response to a localized shear transformation and the local friction angle that characterizes the Mohr-Coulomb condition.

Based on this picture, we proposed a scenario in which spatial correlations of the scattered plastic activity that takes place before the yield point, and the spatial structure of the permanent failure planes that can be observed beyond the yield point are explicitly calculated as a function of these simple ingredients, but occur at different directions. 
The discrepancy between the morphology of permanent bands and transient correlations is in full agreement with Le Bouil \emph{et al.} \cite{le2014emergence}, who proposed a scenario based on the phase-coexistence between  correlated  mini bursts and persistent shearing bands. 
Within this picture, the flowing, post-yielding state is intrinsically different from the states explored at small strains, and the yield transition appears similar to a first order, discontinuous transition.

The failure mechanism we have proposed in Sec.~\ref{sec:MFSlipLineAnalysis} can be interpreted as the maximal instability of uniform lines of slip.
Localization of deformation in this form is favorable, as no further stress is built up in the surrounding medium \cite{martens2012spontaneous}.
Furthermore, the intense shearing effectively lowers the yielding strength inside the localization band (see Sec.~\ref{sec:MFSlipLineAnalysis}) giving rise to a weakening process. 
The maximally unstable modes are dictated by both the elastic interactions \cite{tyukodi2016depinning} and the Mohr-Coulomb plasticity criterion (more specifically local friction) and may be viewed as a major source of mechanical instability \cite{rudnicki1975conditions}.
It is noteworthy that the incurred damage is an ingredient of the model through the failure criterion, and does not enter as an extra material parameter. 

The applicability of our analysis to a real granular medium may be questioned on two accounts: firstly, one may argue that a real granular medium is not described by an linear elastic continuum, due to the existence of force chains and micro plasticity in the form of contact breaking and formation. 
Second, the very existence of a Mohr-Coulomb criterion at the local scale is a rather arbitrary assumption, although in general a pressure sensitivity of the shear transformation is expected. 
Therefore, we have numerically tested our proposition by building
a lattice-based model that incorporates rigorously, if perhaps artificially, these ingredients. 
The results of these simulations are in good agreement with the theoretical expectations, and, when compared to experiments, can lead to a prediction of an effective \emph{local} friction coefficient.

In addition to confirming the theoretical analysis, the simulations offer the possibility to perform an analysis of the macroscopic failure envelope.
This analysis, performed in in Sec.~\ref{sec:MCFailureEnvelope} yields an orientation, labeled as $\theta_\text{MC}$, that is incompatible with $\theta_\text{sh}$, the shear inclination as derived in Sec.~\ref{sec:bandStruct}.
While the former is purely based on the measured bulk stresses, the later arises from kinematic considerations only.
This discrepancy is in contradiction with the Mohr-Coulomb phenomenology at the global scales.
Similar experimental observations were made (see \cite{bardet1990comprehensive} and the references herein) leading to \emph{ad-hoc} remedies that incorporate extra material parameters, \emph{i.e.} the dilatancy angle \cite{vardoulakis1980shear}, into the theoretical framework.   
%
\appendix
\section{Derivation Of The Oseen Tensor For A Compressible Medium}\label{sec:stz}
The force balance equation in a continuum reads
\begin{equation}
	\label{eq:forceBalance}
	\partial_\alpha \sigma_{\alpha\beta}+f_\beta=0,
\end{equation}
where $\sigma_{\alpha\beta}$ is the stress tensor and $f_\beta$ is the applied force per unit volume. 
Rewriting the above equation in the $q$-space, it follows that
\begin{equation}
	\label{eq:forceBalanceQ}
	iq_\alpha~\bar\sigma_{\alpha\beta}+\bar f_\beta=0,
\end{equation}
where $q_\alpha$ is the (longitudinal) wave vector. 
Here the bars denote the transformed fields.

In a linear isotropic homogeneous elastic medium
\begin{equation}
	\label{eq:elasticity}
	\bar\sigma_{\alpha\beta}=\lambda\bar\epsilon_{\gamma\gamma}\delta_{\alpha\beta}+2\mu\bar\epsilon_{\alpha\beta},
\end{equation}
with $\lambda$ and $\mu$ being the Lam\'{e} constants and the strain tensor $\bar\epsilon_{\alpha\beta}$ is
\begin{equation}
	\label{eq:strain}
	\bar\epsilon_{\alpha\beta}=\frac{i}{2}(q_\alpha~\bar u_\beta+q_\beta~\bar u_\alpha),
\end{equation}
where $\bar u_\alpha$ denotes the displacement vector.
Inserting Eq.~\ref{eq:elasticity} into Eq.~\ref{eq:forceBalanceQ} and using Eq.~\ref{eq:strain}, it follows that
\begin{eqnarray}
	\label{eq:lame}
	O_{\alpha\beta}^{-1}\bar u_\beta &=&\bar f_\beta, \nonumber \\
	O_{\alpha\beta}^{-1} &=& (\lambda+\mu)q_\alpha q_\beta+\mu q^2 \delta_{\alpha\beta},
\end{eqnarray}
where the Oseen tensor $O_{\alpha\beta}$ represents a general Green's function of the problem and $q^2=q_\alpha q_\alpha$.
The above tensorial form can be recast in two dimensions as the following
\begin{equation}
	\label{eq:lameAlter1}
	O_{\alpha\beta}^{-1}=(\lambda+2\mu)q_\alpha q_\beta+\mu q^\perp_\alpha q^\perp_\beta,
\end{equation}
or
\begin{equation}
	\label{eq:lameAlter2}
	q^4O_{\alpha\beta}=(\lambda+2\mu)^{-1}q_\alpha q_\beta+\mu^{-1} q^\perp_\alpha 	q^\perp_\beta,
\end{equation}
where $q^\perp_\alpha$ is the transverse wave vector.
The inversion was performed on the grounds that longitudinal and transverse waves emerge as the eigenmodes of the linear operator.
 
Let the bulk modulus be $K=\lambda+\mu$ in two dimensions. Therefore,
\begin{equation}
	\label{eq:oseen}
	O_{\alpha\beta}=\frac{1}{\mu q^2}(\delta_{\alpha\beta}-\frac{\hat{q}_\alpha \hat{q}_\beta}{1+\frac{\mu}{K}}),
\end{equation}
with $\hat{q}_\alpha=\frac{q_\alpha}{q}$ and $\delta_{\alpha\beta}$ being the Kronecker delta.
An effective source due to a localized distortion is expressed as
\begin{eqnarray}
	\label{eq:effectiveReal}
	f_\beta &=& -\partial_\alpha(2\mu\epsilon^\text{stz}_{\alpha\beta}), \nonumber \\
	\epsilon^\text{tz}_{\alpha\beta} &=& \epsilon^*(\delta_{x\alpha}\delta_{x\beta}-\delta_{y\alpha}\delta_{y\beta})a^d\delta(r),
\end{eqnarray}
where $\epsilon^*$ is the total released strain over the local volume $a^d$ and $\delta(...)$ denotes the delta function.
Therefore, 
\begin{eqnarray}
	\label{eq:effectiveFourier}
	\bar f_\beta &=& -iq_\alpha(2\mu\bar\epsilon^\text{stz}_{\alpha\beta}) \nonumber \\
	&=& -i(2\mu\epsilon^*)a^d(q_x\delta_{x\beta}-q_y\delta_{y\beta}).
\end{eqnarray}

Using Eqs.~\ref{eq:oseen} and \ref{eq:effectiveFourier} gives 
\begin{eqnarray}
	\label{eq:displacement}
	\bar u_{\alpha} &=& O_{\alpha\beta}\bar f_{\beta} \nonumber \\ 
	&=& -2i\epsilon^*a^d q^{-2}[(q_x\delta_{\alpha x}-q_y\delta_{\alpha y}) \nonumber \\ &-&(1+\frac{\mu}{K})^{-1}q^{-2}(q^2_x-q^2_y)q_{\alpha}].
\end{eqnarray}
Making use of Eqs.~\ref{eq:elasticity} and \ref{eq:strain} allows for the stress purturbation fields to be determined as follows  
\begin{eqnarray}
	\bar p &=& -K(i q_\alpha\bar u_\alpha) \nonumber \\ 
	&=& -\frac{2\mu\epsilon^*}{1+\frac{\mu}{K}} a^d\text{cos}~2\theta, \nonumber \\
	\bar \sigma &=& i\mu(q_x\bar u_x-q_y\bar u_y) \nonumber \\
	&=& 2\mu\epsilon^*a^d-\frac{2\mu\epsilon^*}{1+\frac{\mu}{K}}a^d\frac{1}{2}(1+\text{cos}~4\theta),
\end{eqnarray}
where $\theta$ is the angle of the wave vector $q_\alpha$. 

For a localized dilation $\epsilon^\text{tz}_{\alpha\beta} = \frac{1}{2}\epsilon_v^*\delta_{\alpha\beta}a^d\delta(r)$ with the released volumetric strain $\epsilon_v^*$ , we obtain
\begin{eqnarray}
	\bar p &=&  -\frac{K\epsilon^*_v}{1+\frac{\mu}{K}} a^d, \nonumber \\
	\bar \sigma &=& \frac{\mu\epsilon_v^*}{1+\frac{\mu}{K}}a^d\text{cos}~2\theta.
\end{eqnarray}

\bibliographystyle{plainnat}
\bibliography{\jobname} 
%
 \end{document}